Preprint Version

# A Bibliometric Perspective of Social Science Scientific Communities of Pakistan and India

Sami Ul Haq [a], Saeed-Ul Hassan [a]

[a] Information Technology University, 346-B, Ferozepur Road, Lahore, Pakistan

**Abstract**
Social network analysis is becoming one of the fundamental topics in graph theory and community detection is one of the important sub fields of social networks. Authors in scientific communities collaborate with each other. These collaborations in research groups are made for various purposes, e.g. to achieve a common goal, for knowledge sharing, improve quality of research etc. Collaboration among authors of a community can be measured via network map. In this study, we will use research publication data from the field of social science to identify collaboration networks among social science research communities of India and Pakistan. We have used Scopus database to extract information of social science journals for both countries India and Pakistan. Study of this data is significant as both countries have common social issues and many of common social values. Research groups of both countries with the same cultural background and having same social issues are more likely to work in related areas. Authors from these countries also work with each other for common social issues. For this study, we have compared data of research publications from the field of social science in India and Pakistan. Based on different collaborations maps among authors and countries like co-authorship network analysis, citation network, source citation network analysis, we have compared collaboration of both communities. Although communities of both countries are facing many common social issues, but still, research communities from both countries are not likely to directly work with each other. Based on collaboration maps, it is identified that collaboration among both countries is weaker as compared to other countries like USA, UK, and Canada etc. Quality of research publications for both countries is also measured in this study and based on these measurements, we have seen performance of both communities. Based on published data of both communities, it was measured that quality of Indian research publication is better as compared to Pakistan. Publications of Indian articles are mostly in Q2 ranks journals, while most of Pakistan's research publications are in Q4 rank journals. Keywords analysis has been done to see common research areas in both communities like poverty, education, the issue of gender, terrorism etc. Despite having many of the common social issues, collaboration among social science research communities of both countries is not strong.

## 1. Introduction

Community is group of users who interact and collaborate with each other to share common interest more than other users within a network (Pornprasit et al, 2022). It is not necessary for users to be on the same demographic location but can also be connected virtually. Communities are of different types and exhibit particular properties like heterogeneous communities, explicit or latent communities, partitioned communities etc (Hassan et al, 2016). To explore these properties, one of the major areas in computer science is social network analysis, and it is gaining more attention day by day in the industry as well as in the scientific research field. Computer scientist are interested in finding different patterns within community to find out information associated with in it (Moosavi, 2017). Communities can be identified by studying network graph generated by these communities. This network graph can be made up on individuals or organizations and each entity is be called as node (Abbe, 2015). Nodes can be connected via edges based on specific relationship among these nodes like common research interest, sharing of knowledge, friendship etc. These communities may be of different types like biological or web communities and each community may also include sub communities. Different sort of analysis can be performed on these networks, e.g. identification of collaboration among nodes, finding similarities and dissimilarities among communities, identification of sub communities within a community etc. Scientists can highlight nodes with similar properties or nodes which communicate more often with each other based upon related properties (Bai, 2017). Similarly, information related to cells, diseases and genes can be studied from biological communities. If we compare two communities, then based on resemblance in characteristics, we can classify these communities as overlapping, heterogeneous or embed with other (Planti'e, 2013). This problem of finding communities from different social networks has been studied since a long time. Information fetched from these communities are then used in different analysis and visualizations to make a better decision. Structural intuition and systematic relation models are one of the fundamental elements for finding useful information out of these social networks. Mathematical models are also very helpful to understand the relationship among group of related nodes of community. Although it is difficult to study graph network but once smaller communities are highlighted, it is easy to understand and fetch information from rest of the graph network (Helal, 2017).

## 1.1 Social Community Structure

Social science communities can be effected via many ways; including cultural values of any nation or country, political system of country, ideology of nation etc. (Zajda, 2015). We can observe behavior of human beings in one community and can co-relate it with other communities as well to see similarities or common values (Meloni, 2015). Social science research communities used to work on social issues and problems of society. If a problem is common across different communities, then social science research communities ought to work on same or related patterns to solve these issues (Clark, 2016). Social communities with same cultural background and facing same problems are more likely to do research in the similar patterns. Based on these facts, we want to see that either research communities from India and Pakistan are collaborating with each other to solve common social problems or not? In this thesis, we will discuss research communities of social science of India and Pakistan. These two countries have similar cultural background, facing more or less similar social problems. These both countries were a part of sub-continent once. So, our hypothesis is that "Social science research communities of both countries will be working and collaborating with each other to solve common social issues." We have listed different parameters to identify collaboration of both communities.

In this study, we are using publication data of social science domain for comparing social science research communities of India and Pakistan. We will be identifying collaboration among both countries and comparing communities of both countries based on different factors.

Here are our main objectives from this study:

1. To envisage collaboration network pattern of research communities from both countries.
2. To explore the areas of research focus of both communities in field of Social Science.
3. To identify citation pattern of social science research communities to see how both countries are interacting with each other.
4. To identify the semantic structure of these communities with respect to bibliographic coupling.
5. To identify quality of publications of both communities and observe trends of research publication in different journals.

We will be validating our hypothesis based on above listed objectives and then conclude results. Apparently both communities have resemblance in number of factors like social issues, problems faced by both communities and cultural background etc. But based on above listed points, we shall analyze whether research communities in each country are also same or not. We will see research collaboration patterns and citation patterns of both communities along with quality of research publication of social science research communities.

## 2. Literature Review

### 2.1 Social Back ground of India and Pakistan

Social problems occur due to pathological conditions. Pakistan and India were a part of sub-continent with the same cultural background. There are similarities between both communities; cultural values are same with respect to folk culture and history. Many of the social issues in both countries are also same (eeberg, 2017). There are certain issues faced by both countries i.e. women empowerment, poverty, influence of political parties, education etc. There are many other social issues as well which are different in both countries (Makino, 1998). India and Pakistan both are facing social issues out of group solidarity and homogeneity. Poverty is a common social issue, which was examined in both countries by three different categories in order to combat it; requirement to cover the hunger and shelter, better health needs, minimum amount of food, housing, clothing and education (Ram, 1997).

India is placed in the list of ten poorest countries by World Development Report in 1981, below than Pakistan. To eradicate this problem, '*nationalization*' was adopted in 1967. 20-Point Program was propounded by Indra Ghandi in 1975, for reducing poverty and for the elevation of weaker sections (Ram, 1997). It was for controlling inflation, giving incentive to production, welfare of rural population, lending help to urban middle class and controlling economic and social crimes. Majority of the work has been done to describe poverty in Pakistan without being seriously concentrating on the eradication of poverty (Gorodnichenko, 2015). In 1981, from 84 million to 149 million in 2004, the population of Pakistan has been increased drastically. Another exploration is about the structural adjustment policies and their influence in lessening the poverty has been examined especially in two periods i.e. 1984-85 and 1987-88 (Amjad, 1997). The increased health care centres decreases the line of poverty. The Government of Pakistan initiated a Social Action Program (SAP) to combat this trouble in 1992 by improving education, healthcare, nutrition, family planning, sanitation and supply of water (Bader, 2015). A report from Asian development bank explains the strengthening strategies of cash transfer by Zakat system (Ahmed, 2006). Some researchers also examined unemployment as one of the major problems of India as well as of Pakistan (Shahbaz, 2015). On one hand where India is coping with the sanitary problems, it is also facing blood trenching problem of unemployment (Chand, 1987). The policy planners have brought this issue in five-year program with other parameters to achieve a 3.0% growth in employment. The Uttar Pardesh government has also taken several innovative steps to envisage employment. It is to give the land called 'Bhoomi Sena'to the farmers to cultivate it by giving loans (Chand, 1987). A point to note here is; several employment schemes have been launched during election times, but due to the outrage of opposite parties, these schemes have been failed to work properly. Only public-sector jobs have been on the rise. Schemes for eradicating unemployment, which have been introduced by federal and provincial government under Annual development programs in Pakistan (Chand, 1987). Tehsil Municipal authorities, under provincial authorities has started small scale projects, which give short term employments.it also tells us about different development programs e.g Tameer-e-Watan. Economic activities and employment have been generated through Khushhaal Pakistan in the rural areas. Many other programs and skill trainings have been introduced for the people (Baxi, 2006).

## 2.2 Collaboration among communities

From many years, communities of all kinds are working and collaborating in order to improve overall performance. Similarly, social science communities are also working, which directly have an impact on different areas ranging from economic growth to the change of political policies (region, 2015). This social science work also empowers other communities to participate actively in development (Luo, 2015). The main objective of this study is to find out interaction between different actors of the society. There are different areas of social sciences; education, political science, economic development, sociology, History, Law etc. All areas are equally important, which participate in community development leading towards impact on society (Boulianne, 2015). Scientific communities are producing a lot of data in the form of publications all around the globe. This bibliographic data is available at different sources for comprehensive data analysis (Huang, 2014). This analysis can be done by developing different maps. Construction of maps based on bibliographic data is becoming important from last few decades to study community maps and measure research performance and quality (Windsor, 2013). There are different types of bibliographic maps, providing information related to research communities i.e. collaboration pattern, research topic identification, co citations etc. Scientific communities collaborate with each other in order to improve research performance (Park I. a., 2015). There are many complex projects which are very difficult to handle by one researcher (Yan E. a., 2012). Most of the scientists look for the collaboration to achieve common goals. There are also many other factors; resource sharing, knowledge sharing, idea generation and time saving to improve overall performance. Trend of collaborations among scientific research communities is increasing depending upon the nature of projects. In communities, one can find collaboration network of different authors to see how they work and collaborate with each other. In this research collaboration, role of specific University is also important, because collaboration among two institutes, having same area of interest, will be higher (Badar K. a., 2013). For finding relation between actors, a technique of Logistics regression analysis (LRA) is used for probability co-authorship. KNN can also be used for this factor. This factor is kept in mind that degree of centrality is one. SVM and K-NN are also suggested to use for validating machine model which will help in clustering method (Sampaio, 2017). Along with this, proximity matrix is also being used (Eck, 2011) and this is helpful in measuring link strength of a node within network graph. Other than that, this research also portraits demographic data and people on same academic positions as well. In this paper (Badar K. a., 2014), three major techniques are used including, degree centrality, closeness and betweenness centrality. There are three important results discussed based on these factors and it is observed that relationship among authors in public sectors is weaker than the authors in private sectors (Etzkowitz, 2002). For Data fetching, CV of each author is being used and separated on the basis of first name and last name (Kretschmer, 2007). Three important attributes are kept in mind along with other parameters e.g. age, sector and gender.

## 2.3 Collaboration among Authors

Authors from scientific communities collaborate with each other in order to improve research performance. There are many complex projects which are very difficult to handle by one researcher. To improve overall performance, authors do collaboration with other authors. Most of the scientists look for collaboration to achieve common goals. There are also many other factors as well e.g resource sharing, knowledge sharing, idea generation and time saving to improve overall performance (Laudel, 2001). Trend of collaboration among scientific research communities is increasing depending upon nature of projects (Guo, 2005). In research communities, we can find collaboration network of different authors to see how they work and collaborate with each other. In network of research collaboration, role of specific University is also important to consider, because collaboration among institutes having same area of interest will be higher as compared to other institutes (He, 2009).

There are a number of similarities between India and Pakistan for example, cultural values, geographical area and social issues etc. These common social issues varies from poverty to gender discrimination and education to problems of food. Although both countries are facing separate social issues as well but based on common social issues, we want to observe structure of social science research communities in both countries. Social science research communities work and collaborate with each other to solve different problems. Similarly, scientific research communities also collaborate with each other for different research problems to improve research performance, reduce cost and for resource sharing etc. These collaborations among different actors can be studies in network graph using multiple techniques like clustering, Logistics regression analysis and centrality measurements. In this study, we will be using these techniques to identify collaboration patterns of social science research communities of India and Pakistan.

## 3. Data and Methodology

### Data

In this paper, we have selected the data of social science research publications of Pakistan and India. For the extraction of data, Scopus database is used and research publication's data from social science fields is being fetched. One limitation in Scopus database is; user can extract maximum two thousand records at a time via query or from graphical user interface. We selected data based on different range of the year. For this purpose, data was divided into different chunks of years and after that we combined all data and make one single file for each country. This publication data consist of around 41 columns having variety of information like authors, title, source title, affiliation and affiliation of authors etc. Based on these columns, we extracted different information from both countries.

In previous chapter, we have discussed that Pakistan and India are facing related social issues. So based on our hypothesis,  want to see how research communities in both countries are working with respect to collaboration for problems, area of research focus, citation of work and bibliographic coupling. Another strong motivation is that, there has not been any comprehensive analysis done before in social science research communities of India and Pakistan in accordance with convergence, divergence, and evolution of communities. In this thesis, our major focus is to study collaboration pattern among research communities, co-authorship network analysis, co-citation analysis, identification of major research areas leading towards comparison of both communities. Analysis can be done on these nodes on maps, based on Author, Journal, Citation of a journal, citation of an author node, key words from a specific journal and Co-occurrence of words in abstract (Rivellini, 2006). Similarly, maps can be constructed for different countries or departments from publication data. From these maps, we can identify and fetch our required information of communities. These patterns help in identifying new research fields, list of authors which can collaborate in near future, authors with same area of interest and list of research areas for a specific author or any department of any country.

## 3.1 Metrics Used for collaboration pattern

For comparing both communities, we have used following metrics:

1. Co Authorship Network of Pakistan and India's research communities
2. Citation source and network analysis
3. Publication Quality
4. Keywords co-occurrence analysis

These metrics will help in identifying patterns within community and relations of research community in both countries. Co-authorship network gives information of authors interacting and collaborating with other authors with same area of interest. Co-authorship analysis is also an important pattern to identify collaboration among communities. In research communities, authors also work with each other for common work or publication. These research collaborations can be measured on nodes of authors, departments, universities, countries etc. So, this co-authorship network can be constructed on nodes of authors, nodes of countries or nodes of universities where authors collaborate with each other. From co-authorship network, we can measure role of an actor in research community as authors can exchange knowledge, experience, resources etc. for common work. Research communities are bind with each other in number of publications they have authored jointly. Linkage between research departments, institutes, countries can be identified with co-authorship analysis (Kazi, 2017).

Co-citation analysis is one of the most commonly used analysis in research communities to see relationship among two publications. If any Publication cites two other publications then these two publications are known as co-cited. For example, if publication A is using reference of publication B and C for any purpose, then these publications i.e. B and C are co-cited. Here the important point is, relationship defined is based on citation of parent paper. More number of common citations in parent publications, the stronger relationship will be among those publications. Another way to measure relationship among publications is bibliographic coupling. When two publications A and B cite another common publication C, then these A and B publications are known as bibliographic coupled. This is slightly different method as compared to co-citation. In bibliographic coupling, parent publications have a relationship because these publications may be working on related work or related area. That's why they are citing a common publication. Keywords can be extracted from author's specified words along with abstract and title. Based on these keywords, we can identify major areas or topics in research communities being discussed. Major focus is usually co-occurrence of these words in research publications. This can also be used in topical network analysis.

Authors of a research publication can be a part of different communities. Based on their affiliation with university or community, performance can be measured. Author's collaboration pattern can be measured based on the sum of link weights and Co authorship collaborating value (Kazi, 2017). Similarly, scientific cooperation relationship graph, scientific cooperation relationship maps also identify these collaborations (Jiang, 2008). This pattern can be identified based on author, university and department as well. To identify topical interaction in community, it is necessary to study overall collaboration network within community (Eck, 2011).

## 3.2 Methods and Techniques

This study is based on the data of research publications retrieved from Scopus database. This data set includes research publications of all kinds in social science and in its sub domains. Information of research publications from 1947 till 2016 have been fetched from this data base. Columns of the tables include information about authors, affiliations, title of publication, source of publication, year, Abstract, author defined keyword, references, conference information etc. For evaluation purpose, two sets of data have been prepared. One of the data sets consists of research publication with at least one author from Pakistan and other include publication's data with at least one author from India. Our aim in this study is to compare the results of co-authorship analysis, bibliographic coupling etc.

We have used Biexel, which is used for Social Network Analysis and it is freely available. Majority of comparison has been done with VOSViewer which is also free and used to analyse bibliographic data (van Eck, 2013). We have selected link strength as key component for analysis purpose. Different visualizations are available in this tool including network visualization, overlay visualization, and density visualization (Van Eck, 2009). We have selected network visualization to indicate closest clusters and their relationship based on link strength (Bashan, 2012). The strength of a social link depends on the frequency and the length of social interactions. For measuring link strength, nodes from clusters are identified and measure the interaction between them.

More interactions will lead to stronger link strength. It is also an important factor to measure multiple types of clusters in network. It is a good parameter for modularity data (Graves, 1999). Other than that, Normalization is used for analysis of the data. Association strength is used to normalize the link strength of links between nodes.

For the construction of maps, clustering method (Waltman, 2010) is being used. Clustering is grouping of different nodes in such a way that nodes of same attributes are placed in one cluster and nodes with different attributes are placed in different clusters. Nodes which are similar to each other are identified easily. This technique also helps in studying relationship among different nodes.

Co-authorship gives information about common work done for a specific research publication (Yan E. a., 2012). It can be measured among different authors of a department, authors of different universities and authors among countries (Bazm, 2016). In this study, we will measure co-authorship among authors of countries (Kwon, 2012). Co-authorship network is considered to be an undirected graph. Nodes are represented as countries, while links are indicating collaboration among nodes with respect to number of papers both nodes co-authored (Ta{\c{s}}k{\i}n, 2015). Network map is undirected weighted co-authorship graph, in which G is a pair G = (N, m), where N = {1……. n} is a finite set of nodes (where nodes are countries) and m is a n*n matrix, in which mij represents the weighted relation (number of papers written together) between country i and j, with mij = mji. Consider the graph G = (N,m) where N is number of Nodes and m is number of edges. In case of directed graph, mij =1 if there is an edge between 2 nodes I and j. But in our case, our graph is un directed so we assume that if m*ij* or m*ji* = 1 if either if I and j are connected.  Total number of links are indication of number of countries that are co-authors. Therefore, the total number of links indicate the total number of co-authorship relations existing in the network. Number of links between two nodes i and j can be denoted as cij. Strength between two nodes is represented as following.

$$Sij = \frac{2mcij}{CiCj} \quad (1)$$

Where S*ij* is total number of links in node i and m total link strength of a given network. Degree of centrality against these nodes are also calculated. The degree of centrality of vertex v is simply given by number of edges incident upon it. It is also measured by following formula (Fatt, 2010):

$$CD\ (v) = \frac{\deg(v)}{n-1} \tag{2}$$

Co-occurrence analysis of words usually done for visulization of relationship among topics and words defined by authors. This is also used for visulizing relationships among people, organization with research material. For keyword, Co-occur term used to describe when idea or specific word exists in two research material. For example, if a keyword A appears in research publlication B and C then the keyword A is co-occurred with respect to both publications. Using text mining methods (Nawaz et al, 2022, Chansai et. al., 2021, Yin et., at., 2020), one can also predict about potential ideas of author and relationship among people or organizations having same area of interest. Publications which are similar sementically, have more co-occurred keywords. Different metrics of co-occurrence also exist like co-citation and co link metrix, which provide information among two document publication. Closeness Centrality can be used for this purpose. This is, average number of steps required to reach all reachable vertex in a graph. It can be calculated as following (Fatt, 2010):

$$Cc(v) = (n-1)/\sum i \neq j\ d(i,j) \tag{3}$$

For co-occurrence of words, pair of documents are identified and a graph is constructed among common words appearing in publications. Let n be the number of co-occurences in the list of vertex having degree N, then n is measured as binomial coefficent of N over 2. This n will increase with the increase of vertex dergree of graph. Number of co-occurrence of a specific word can be calculated as sum of expected co-occurrence in all list. Expected co-occurrence can be calculated as $(N_l-1)*p_l$. P1 is occurrence of word that is selected by null model with specific distribution of data (M{\"u}ller, 2008).

## 3.3 Tools Used

We have used different tools in this study. For data cleaning and removing outliers, we have used R and R Studio. We have used Biexcel and VOSViewer for finding collaboration pattern and for construction of different clusters on data of research publications. For finding citation and quality of work being published by both countries, we have used MS SQL server tool and SQL language is being used for scripting purpose.

Based on metrics defined above, we have constructed different maps of both communities and collaboration pattern among both communities are identified. Similarly, based on constructed clusters, we have analyzed maps of co-authorship network, bibliography data and key-words analysis. Quality of research publication from both countries is also measured.

## 4. Results and Discussion

In this chapter, we have discussed our results based on the data of social science research publications from both communities. This chapter is further divided into different sections based on metrics described.

### 4.1 Co Authorship Network of Pakistan and India's Research communities

To study how authors are participating in social science research communities, different types of co-authorship networks can be studied for example, co-authorship network within one country with respect to departments and collaboration with other countries etc. Here, we are presenting collaboration with other countries.

Let's first talk about co-authorship network of Pakistan's research community. In following picture, it is displayed that how authors of Pakistan from social science research domain are collaborating with authors from other countries. Here each country is displayed as one node. With respect to Co-Authorship network, authors from Pakistan are more actively interacting with USA, UK as displayed in link strength table and also displayed in this picture.

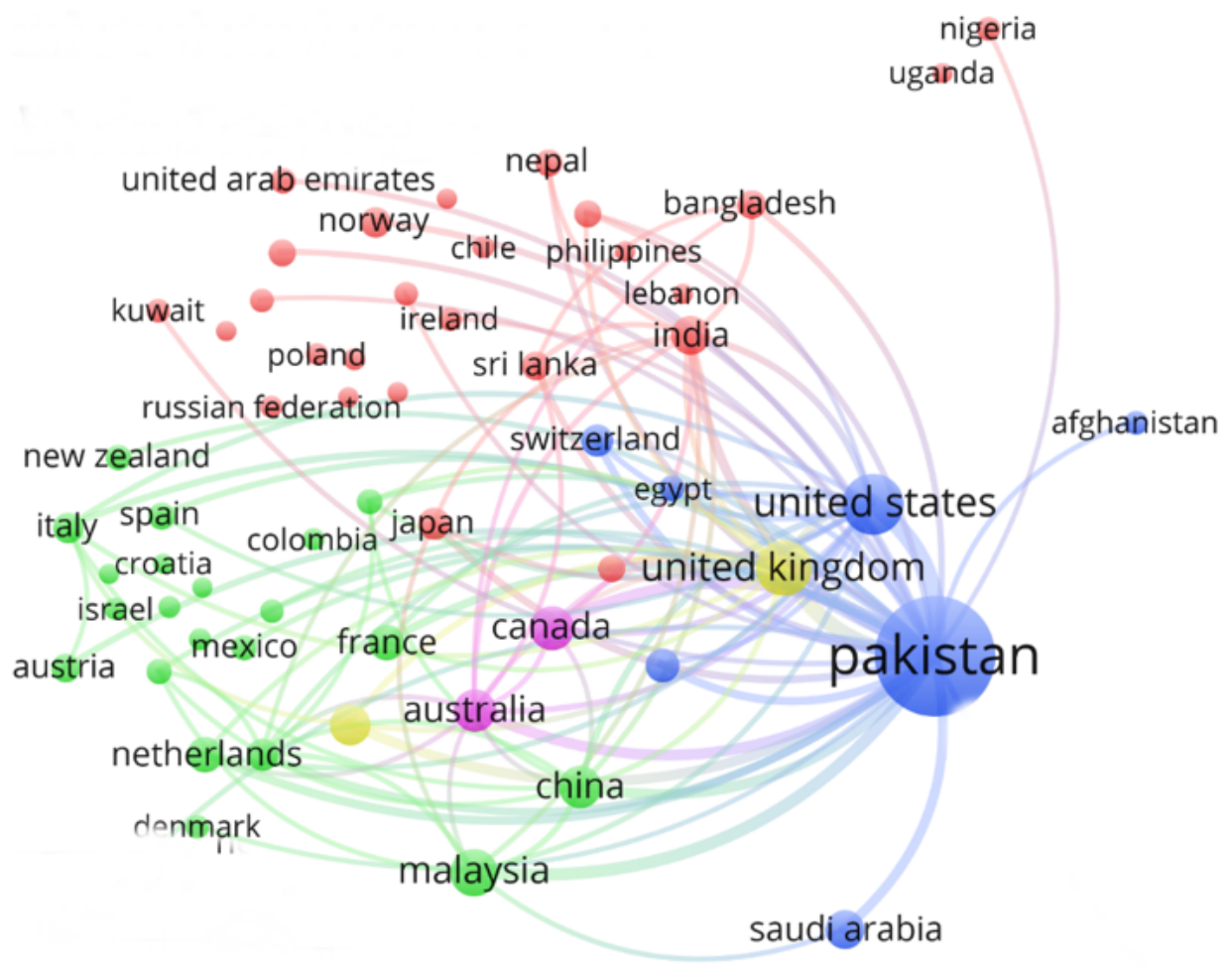

**Figure 1:** Co-Authorship collaboration network of Pakistani authors

**Table 1:** Link Strength table of Co-Authorship collaboration network of Pakistani authors

| **Country** | **Link Strength of Co-Authorship Network** |
|---|---|
| USA | 249 |
| UK | 238 |
| Malaysia | 126 |
| Canada | 101 |
| Australia | 87 |
| China | 81 |
| Germany | 68 |
| Saudi Arabia | 65 |
| India | 62 |
| Netherlands | 45 |

This table is representing link strength of Co-Authorship collaboration network of Pakistani authors with other countries. While talking about map of Pakistan, this is the pictorial representation of Co-Authorship collaboration network of Pakistani authors with other countries. Collaboration among Pakistani and Indian authors, in the field of social science domain is not that strong as compared to collaboration with other countries. Although both countries are facing many common social science issues as mentioned above but research communities are not working with each other to that extend. If we see collaboration from different clusters, then it is obvious that collaboration among authors of some of specific regions is prominent. For example, authors of different European countries like France, Austria, Spain, Italy etc. lies in one cluster as shown in above picture. On the other hand, India, Srilanka and Bangladesh are lying in one cluster. There is a possibility that authors from different regions are collaborating with each other for related social problems faced by these regions. Pakistan, Saudi Arabia, USA, Switzerland lie in one cluster. All of them are from different regions. If we talk about social issues and social values, then all of these countries have different social problems and different values, but depending upon linkage between authors of these countries, collaboration rate in research is high. So we can say that, authors from Pakistan and India are collaborating with other authors depending upon basis of personal linkage. Having same social issues and social background is not a key factor in research collaboration among authors.

Now let’s talk about co-authorship network of Indian authors.

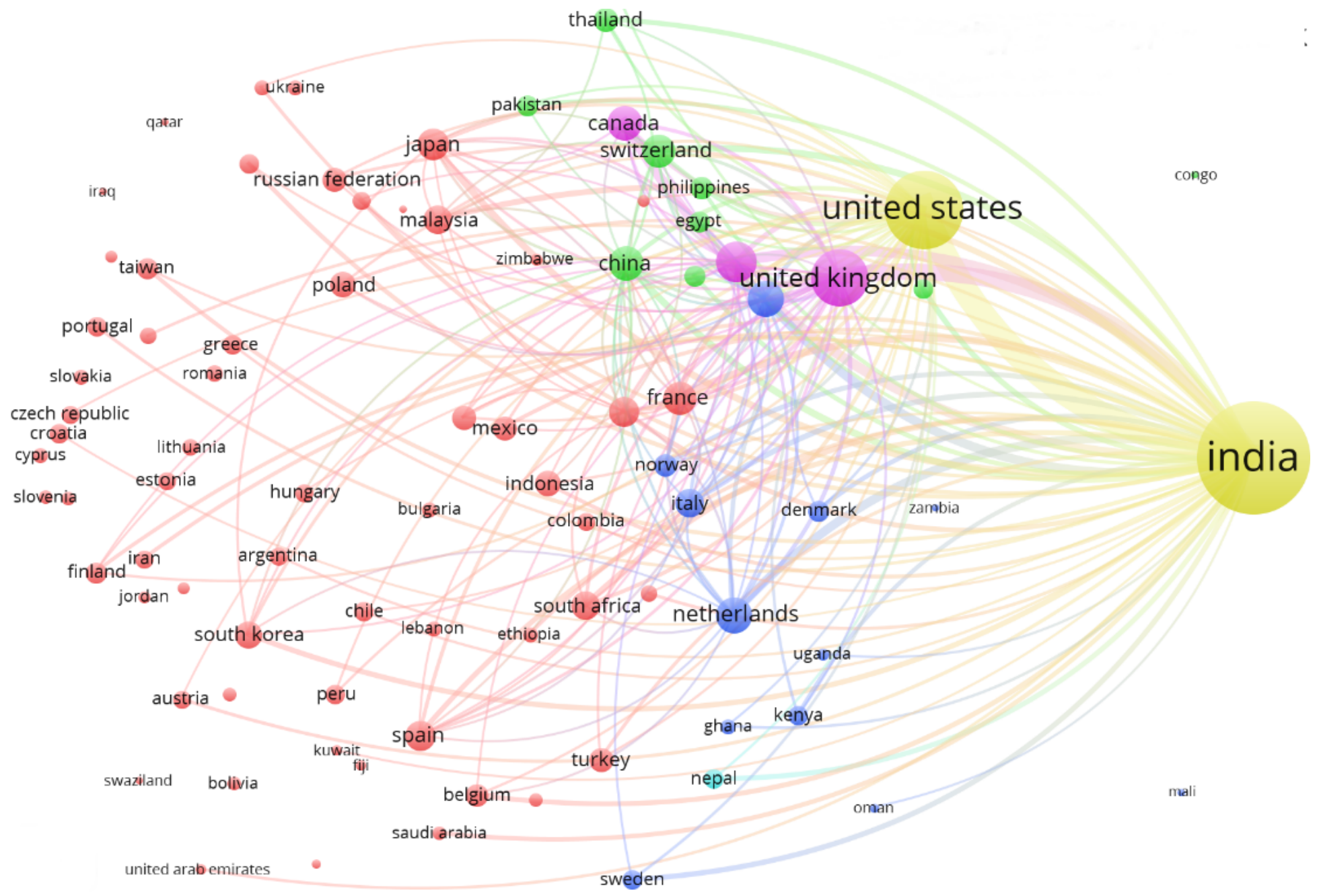

**Figure 2:** Co-Authorship collaboration network of Indian authors

**Table 2:** Link Strength table of Co-Authorship collaboration network of Indian authors

| **Country** | **Link Strength of Co-authorship Network** |
|---|---|
| USA | 2315 |
| UK | 938 |
| Australia | 384 |
| Canada | 359 |
| Germany | 281 |
| Netherlands | 274 |
| France | 192 |
| China | 166 |
| Japan | 166 |
| Netherlands | 45 |

This map is a collaboration network of Indian authors with authors of all other countries. India and Pakistan are collaborating more or less with same countries at different levels. But, here again, it is clear that collaboration with Pakistani authors is very weak as compared to other countries. Most of the time, Indian authors are collaborating with research authors from USA. In above picture, USA and India also exist in same cluster despite of different regions. In above map, collaboration of authors related to specific regions is visible. For example more than 70% of European countries exist in one cluster and authors from these countries are collaborating. Similarly, authors from Asian countries are displayed in one cluster e.g Pakistan, China, and Thailand etc. In collaboration map of India, it is also obvious that collaboration with Muslim countries is not as strong as compare to the other countries. There are different Muslim countries in Asia i.e Pakistan, Saudi Arabia, Indonesia, Malaysia etc, but collaboration of Indian authors with these countries is weak as compared to the other countries.

## 4.2 Citation with respect to Countries

Author's collaboration can also be measured in terms of citing the same work. Citation is usually done for giving credits to other researchers, if their collaboration is being used in current work. Authors working in same domain are more likely to cite research work of other authors with same area. For example, authors working on problem A will like to cite work of those authors who have been involved in solving related problems. To understand citation of authors from Pakistan and India, citation network for both countries has been created. Let’s talk about citation pattern of Pakistan’s authors.

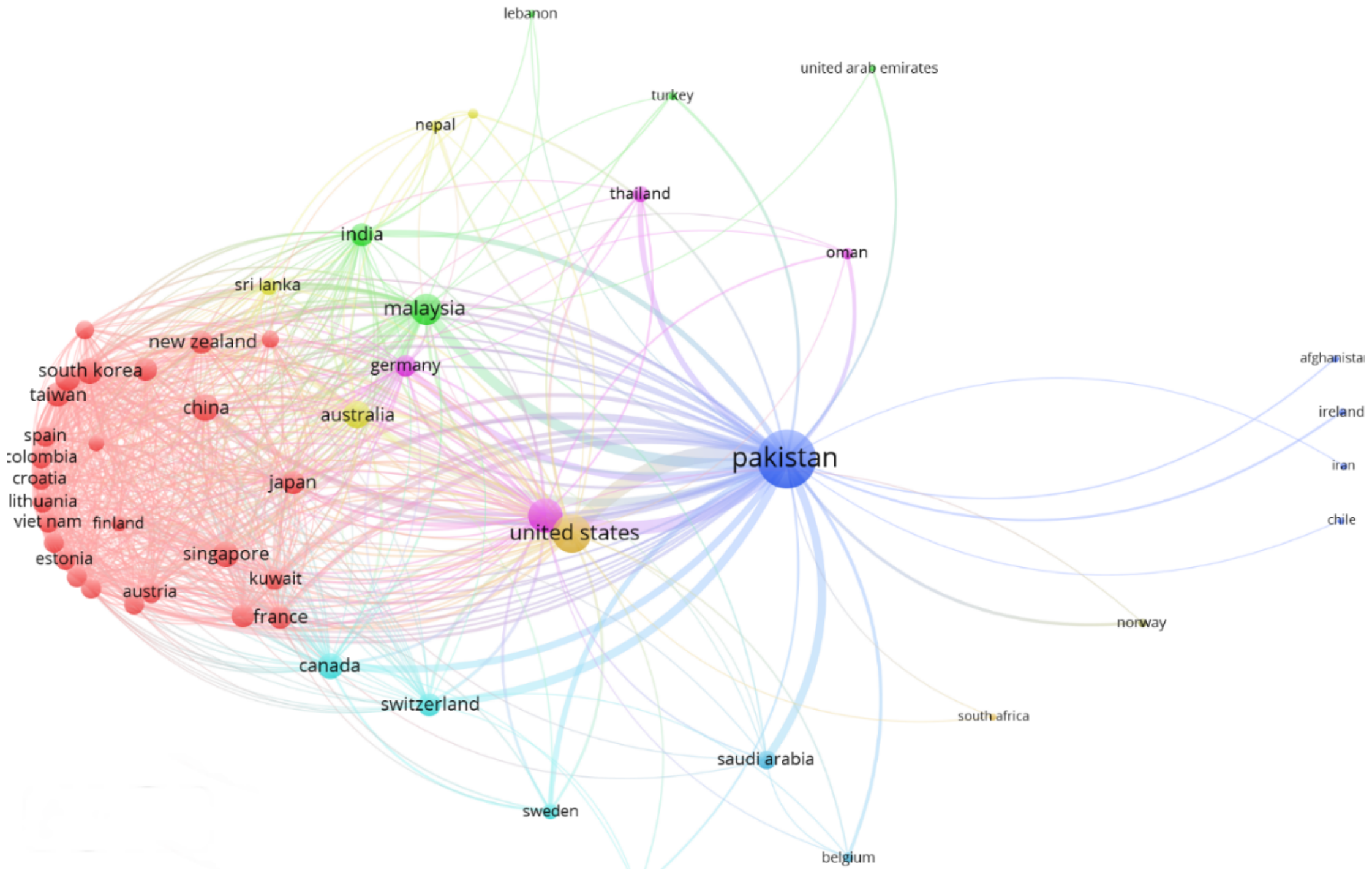

**Figure 3:** Citation network of Pakistani authors

**Table 3:** Link Strength table of Citation network of Pakistani authors

| Country | Link Strength of Citation Network |
|---|---|
| USA | 350 |
| UK | 256 |
| Malaysia | 213 |
| China | 149 |
| Australia | 139 |
| Singapore | 131 |
| Canada | 126 |
| South Africa | 122 |
| Brazil | 120 |
| Taiwan | 120 |

Citation network of Pakistan is similar to the co-authorship network of Pakistan. From this citation network, it is clear that Pakistani authors are not citing research work from Indian authors as compared to other countries. Authors from Pakistan are citing more work done in European countries as seen in the image. Countries collaborating with each other more frequently exist in one cluster. It is also displayed that most of the European countries lie in one cluster. India, Malaysia, United Arab Emirates, Turkey Lie in same cluster. If we see link strength table then it will be clear that most of the developed countries does not lie in same cluster e.g. USA and UK, Australia lies in different clusters. China and Canada also lie in different clusters. As stated earlier that our assumption was; depending upon common social issues, both countries will cite work of each other to find common ground. But in this case, it is not true. Citation pattern is strong with those countries having more collaboration with Pakistani authors. Now let’s have a look on citation pattern of India’s research community.

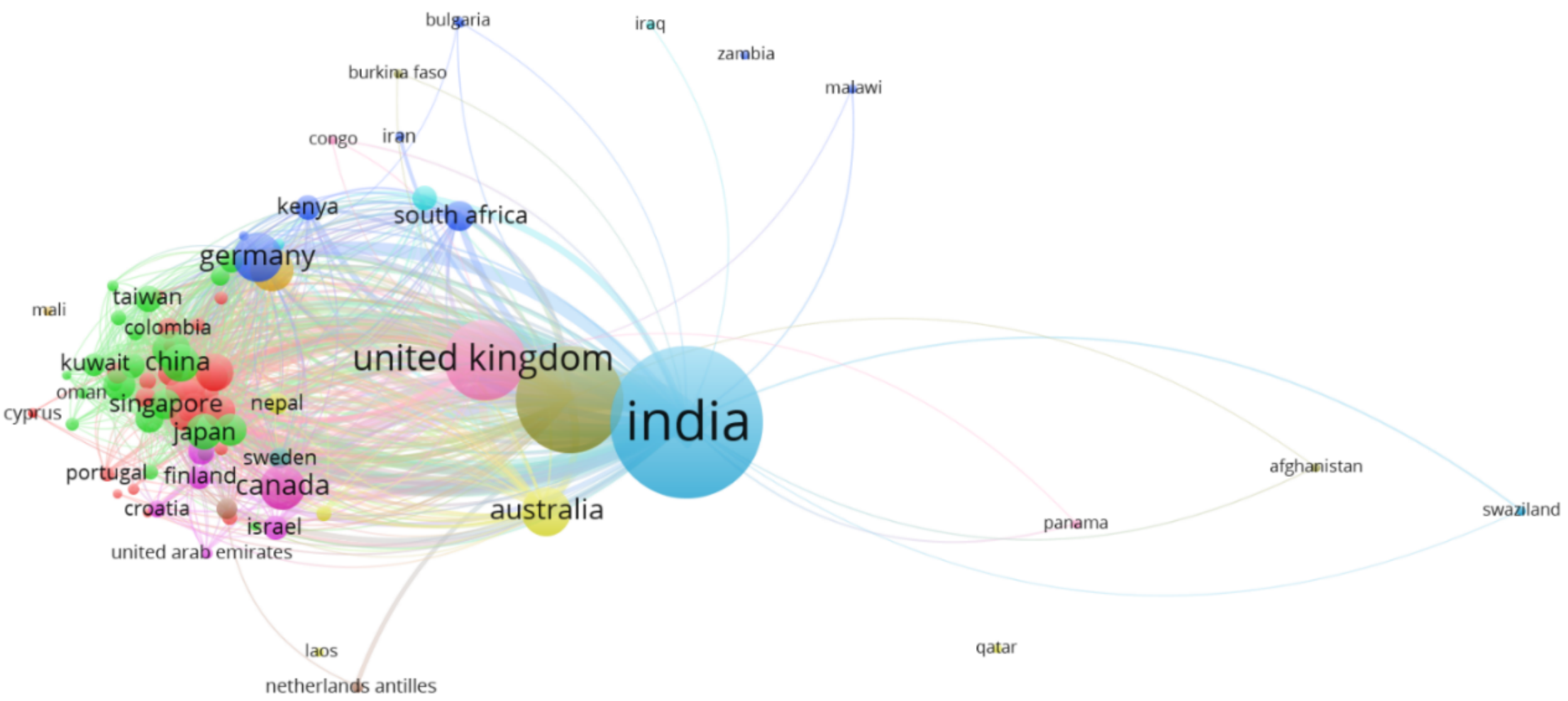

**Figure 4:** Citation network of Indian authors

**Table 4:** Link strength table of Citation network of Indian authors

| **Country** | **Link Strength of Citation Network** |
|---|---|
| USA | 3580 |
| UK | 2094 |
| Australia | 796 |
| Germany | 753 |
| Canada | 655 |
| Netherland | 606 |
| China | 482 |
| Brazil | 453 |
| France | 443 |
| Japan | 425 |

Citation pattern of Indian authors is also related to co-authorship collaboration pattern. In citation network although both countries (India and Pakistan) are citing publication with common top five countries but clusters of these countries are not same. These countries exist in different clusters indicating that researchers from Pakistan and India are not working in same pattern for citing work from other countries.

## 4.3 Source citation network

Source is any conference or journal where some research work has been published. Citation of sources may be used to access one's publication quickly. Sometimes, quality of publication is also determined based on published source and citation from sources.

For example, if research work is published in good journal A and citation is also used in this publication from good journals then it is considered that this research is good. Similarly, if research publication published in some average source B and cited publication's source is also average then this research work is considered to be average. That's why source is important, for publication and citation as well. For this analysis, we have bench marked data and selected at least 20 documents per source.

For Pakistan, we find only 15 connected sources out of 30 qualified sources who published more than 20 documents. And out of these connected sources, 4 sources are from Pakistan.

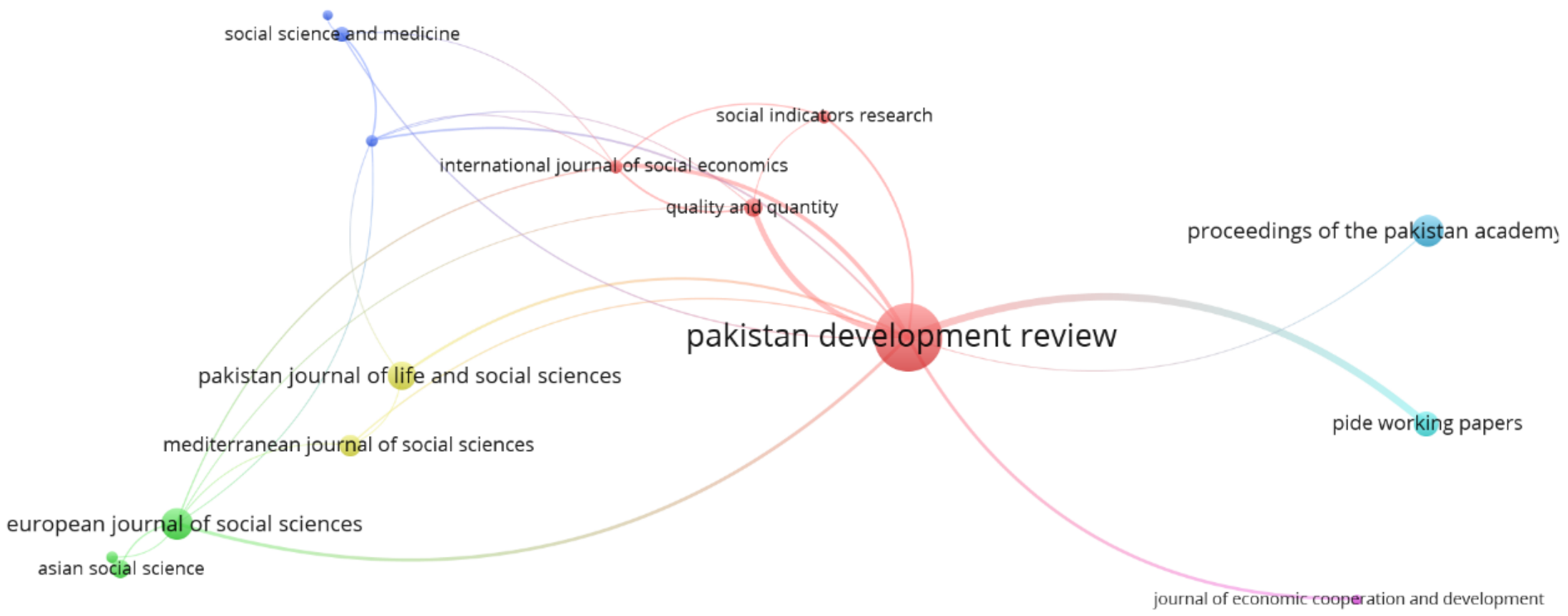

**Figure 5a:** Source Citation network of Pakistan

Here is the chart of Source Citation and Document count from Pakistani research publications.

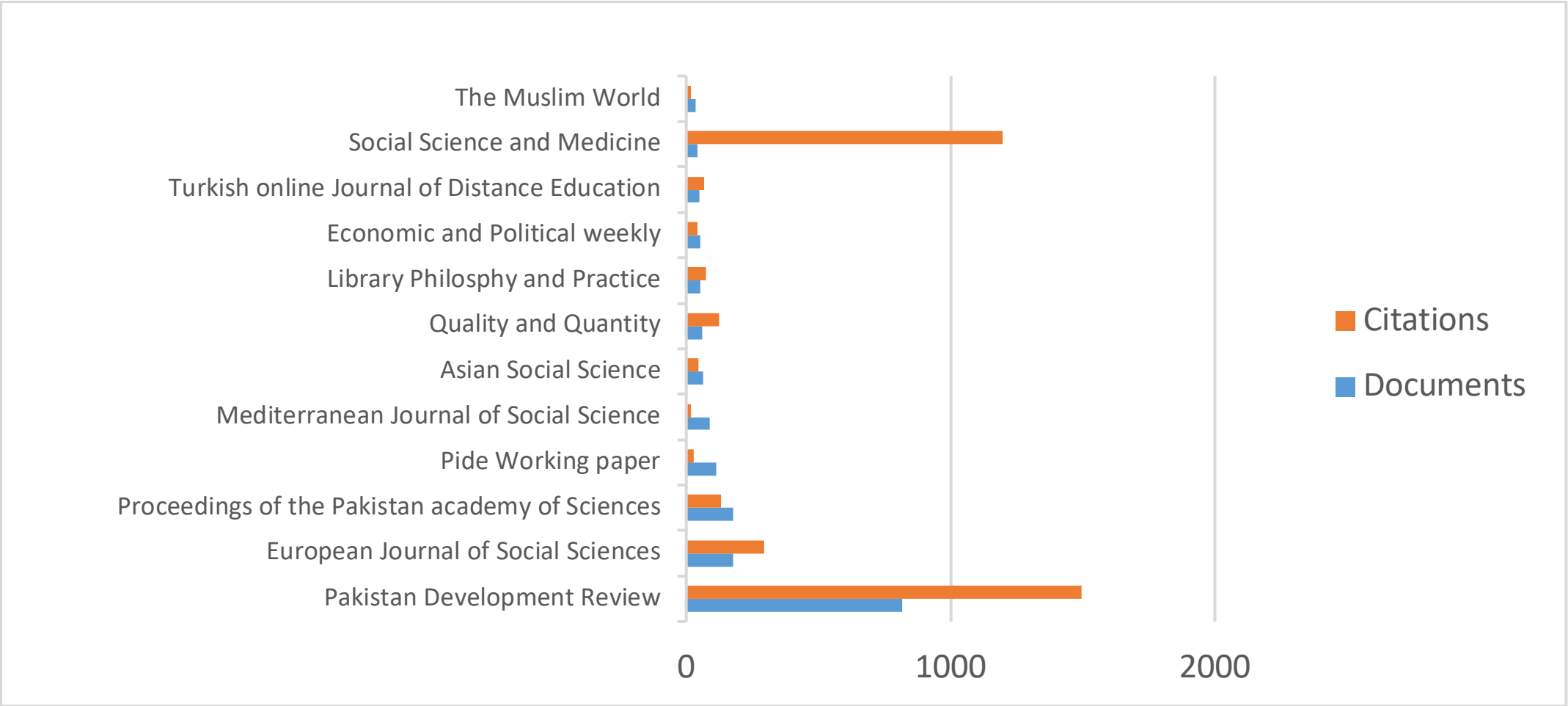

**Figure 5b:** Source Citation and Documents count of Pakistani Research Publications

In above chart, there are some sources for which number of documents are less than citations, for example, Pakistan development review and Social science and medicine etc. have more number of cited documents as compared to number of documents published in these sources from Pakistan's research work. While other sources have more number of documents published as compared to cited documents. Pakistan Development Review is on the top for document publishing from Pakistani authors. Other than international sources, Pakistan have one source from Muslim community as well named "The Muslim world" which is dedicated for publications on Islam, Muslims and current Muslim-Christian relationship. Being in Asian zone, authors from Pakistan are also contributing in Asia specific source that is "Asian Social Science".

Now let's see source network of Indian data.

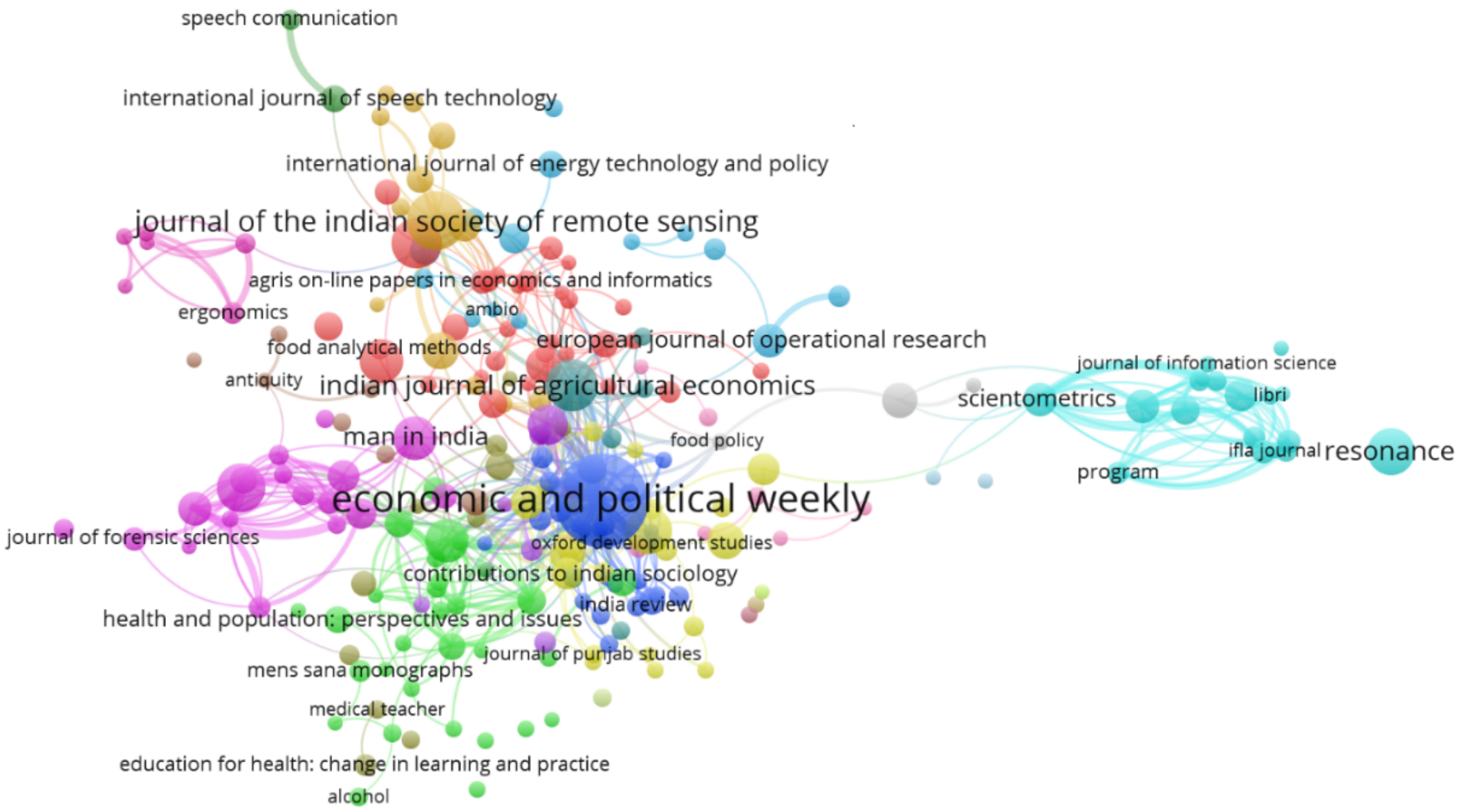

**Figure 6:** Source Citation network of India

Here is the chart of Source Citation and Document count from Indian research publications.

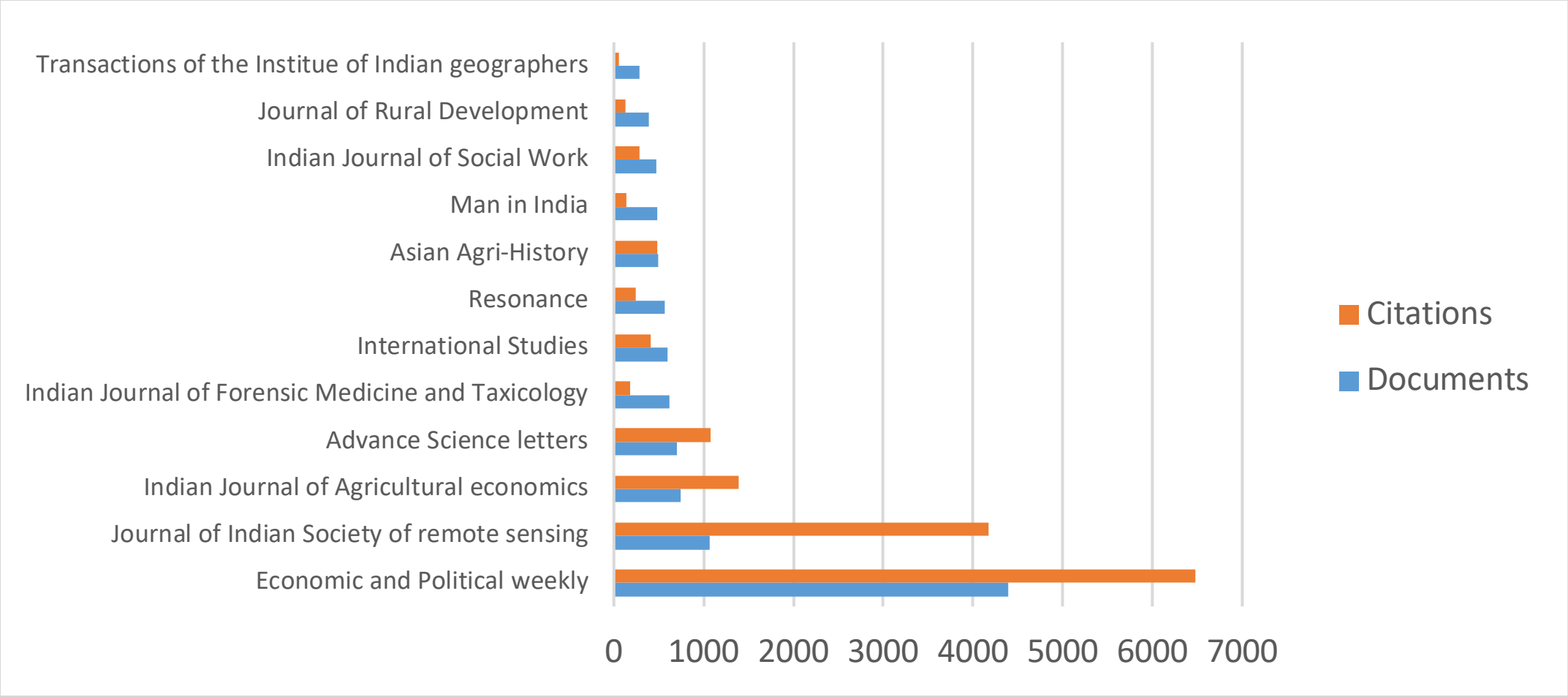

**Figure 6b:** Source Citation and Documents count of Indian Research Publications

From top 10 source documents count of Indian data shows that authors from Indian community are publishing in local sources journals more frequently as compared to international sources. While, from top 10 source documents count of Pakistani data, only 2 local sources are identified for publication. But overall, India has more international sources as compared to Pakistan. Both countries has "Economic and Political weekly" source as one of top documents publishing source. But India has this sources on top indicating that they are more focused towards economic development and concerned about politics. Publisher of this source is also Indian. Based on these results, we can say that Indian authors are also focused on technology field even in social science. And they are doing more research in field of agriculture as compared to Pakistan as there 2 major sources are from Agriculture domain. Our previous hypothesis is also not true here about using common source for common social and culture values as sources of publications and citations are also not common in both countries.

## 4.4 Publication distribution with respect to SJR Quartile

To measure research quality of any research community, we can see how many of research publications are being published in good conferences or in good Journals. Based upon this data, we can predict about quality of any research community. For measuring research quality of both communities, we have compared rank of journal as per SJR quartile. There are four quartiles based on SJR ranking which are Q1, Q2, Q3 and Q4. All of published Journals lies in one of the above category. Journal with Q1 ranks are considered to be of more quality work and then comes Q2, then Q3 and in last Q4.

Here is the chart showing the quartile wise distribution of publications among different Journals.

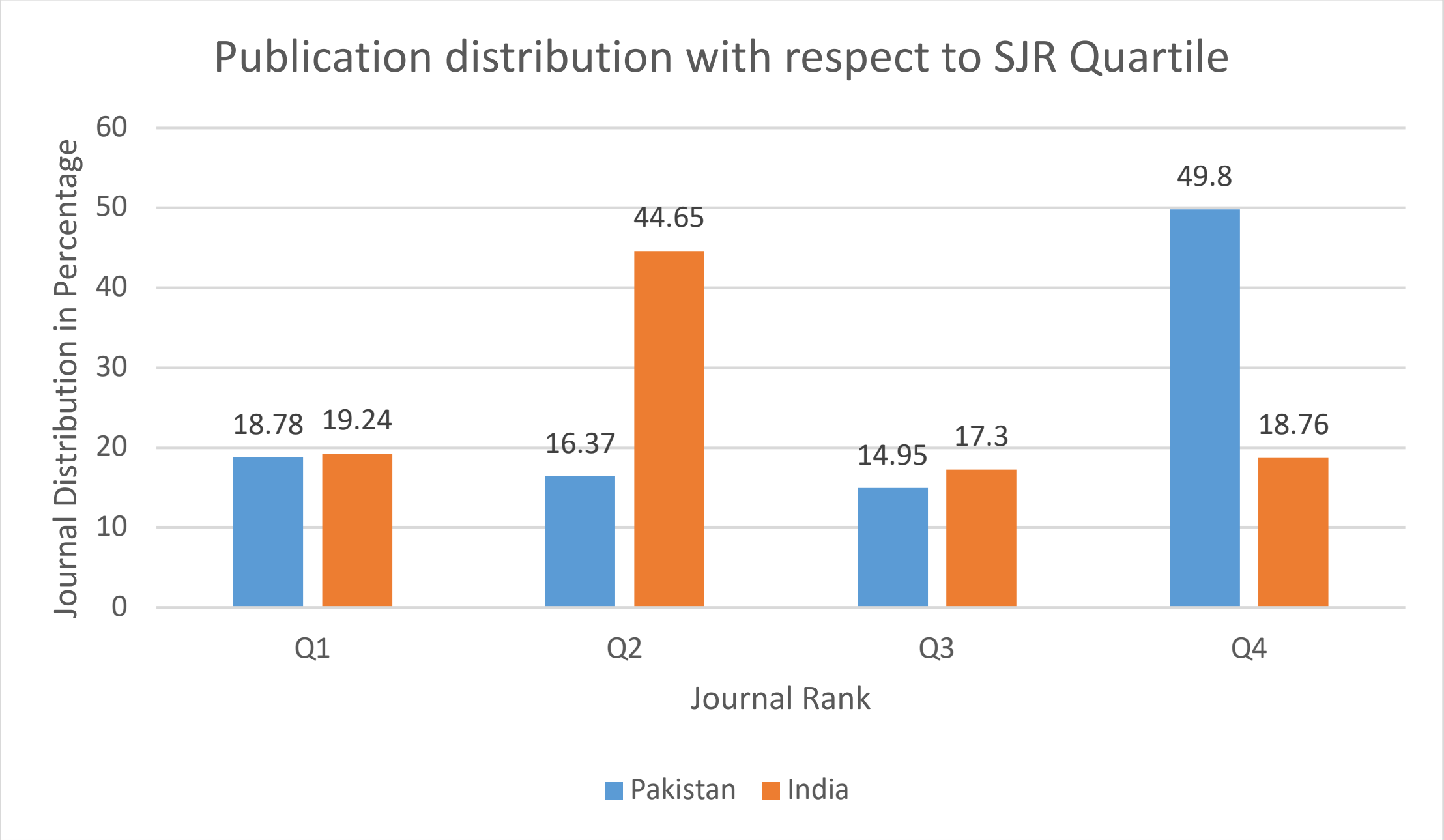

India has higher number of publications in Q1 quartile journals as compared to Pakistan. Pakistan has 18.78% of its publications are in Q1 quartile journals and India has 19.24%. For this quartile, difference is very small e.g. 0.46%. But when we see publications in Q2 quartiles for Pakistan and India, difference is very signification. Pakistan has 16.37% of its publications in Q2 quartile journals while India has 44.65% in this quartile. In Q3 quartile journals, Pakistan has 14.95% and India has 17.3% publications in this quartile journals. In Q4 quartile journals, Pakistan has 49.8% and India has 18.76% of publications in this quartile journals. Based on publications in different quartiles, most of the publications from Pakistan's research community, are in Q4 quartile journals. While, most of Indian's research publications are in Q2 quartile journals.

India owns 2 journals from social science domain which lies in Q1 quartiles. 5 journals in Q2 quartiles are from India. While in Q3 and Q4 quartiles, there are 6 and 5 journals respectively which are from Indian social science journals. While talking about Q1 quartiles journal from Pakistan, none of its journal qualifies for this category from social science domain. Only 1 Journal is in Q2 quartiles that is 'Pakistan Journal of Information Management and Libraries'. Similarly there is only one journal in Q3 quartile from Pakistan research journals and two of its journals are in Q4 quartile journals.

### 4.5 Bibliographic Coupling

This is another metric to see if some of research groups are bibliographic coupled or not. If bibliographic coupling exist among different communities then we can say that these communities are working on same or related area of research. Hence similarity can be established among different research communities. According to definition bibliographic coupling, if two publications Y and Z cites same publications A, B and C then these two publications Y and Z will be bibliographic coupled publications. Number of common publications cited by two publications decides how strength of bibliographic coupling among two publications (Park J. Y., 2013). For example if two publications A and B are citing m number of common publications and other two publications Y and Z are citing m-n number of common publications then bibliographic coupling among A and B will be stronger as compared to Y and Z.
Now let's see bibliographic coupling network of Pakistan's research publications with respect to countries.

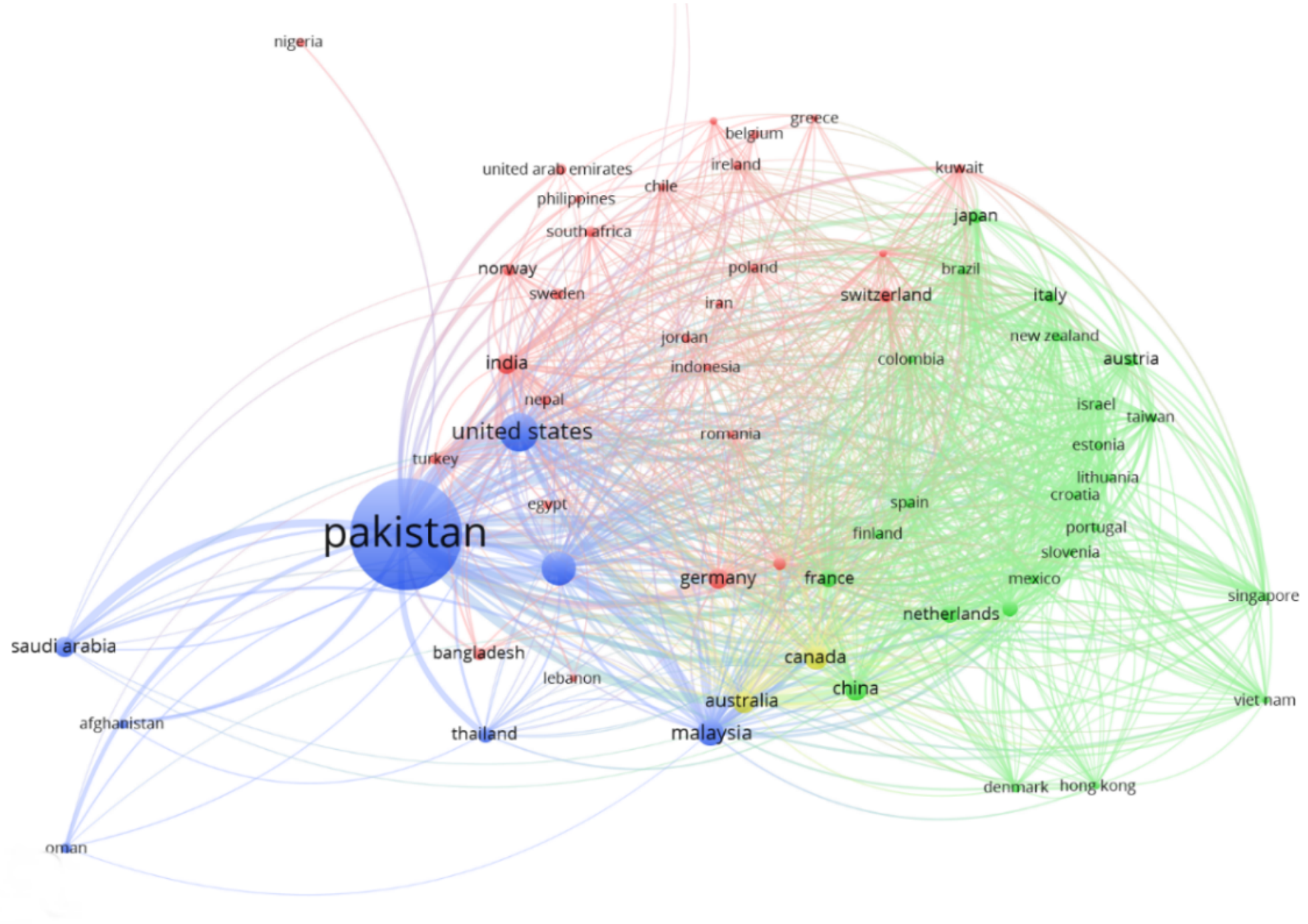

**Figure 7:** Bibliographic coupling network of Pakistan research publications

**Table 5:** Bibliographic Strength of Pakistan research publications

| **Country** | **Bibliographic Strength** |
|---|---|
| USA | 17322 |
| UK | 12849 |
| Malaysia | 9489 |
| Australia | 4891 |
| China | 4778 |
| Canada | 4665 |
| Germany | 3552 |
| India | 3508 |
| Saudi Arabia | 3446 |
| France | 2898 |

According to clustering map and bibliographic strength table, Pakistan is highly coupled with USA with respect to Publications. Means Pakistan's social science research community and USA's research community are citing many of common publications among their work. We can say that both countries are working on related problem. But India comes on 8$^{th}$ number in strength table despite of having same social issues and cultural background. So indication is, although both countries are working on related areas but they are not bibliographic coupled to that extend.

Now let's talk about Bibliographic coupling map of India.

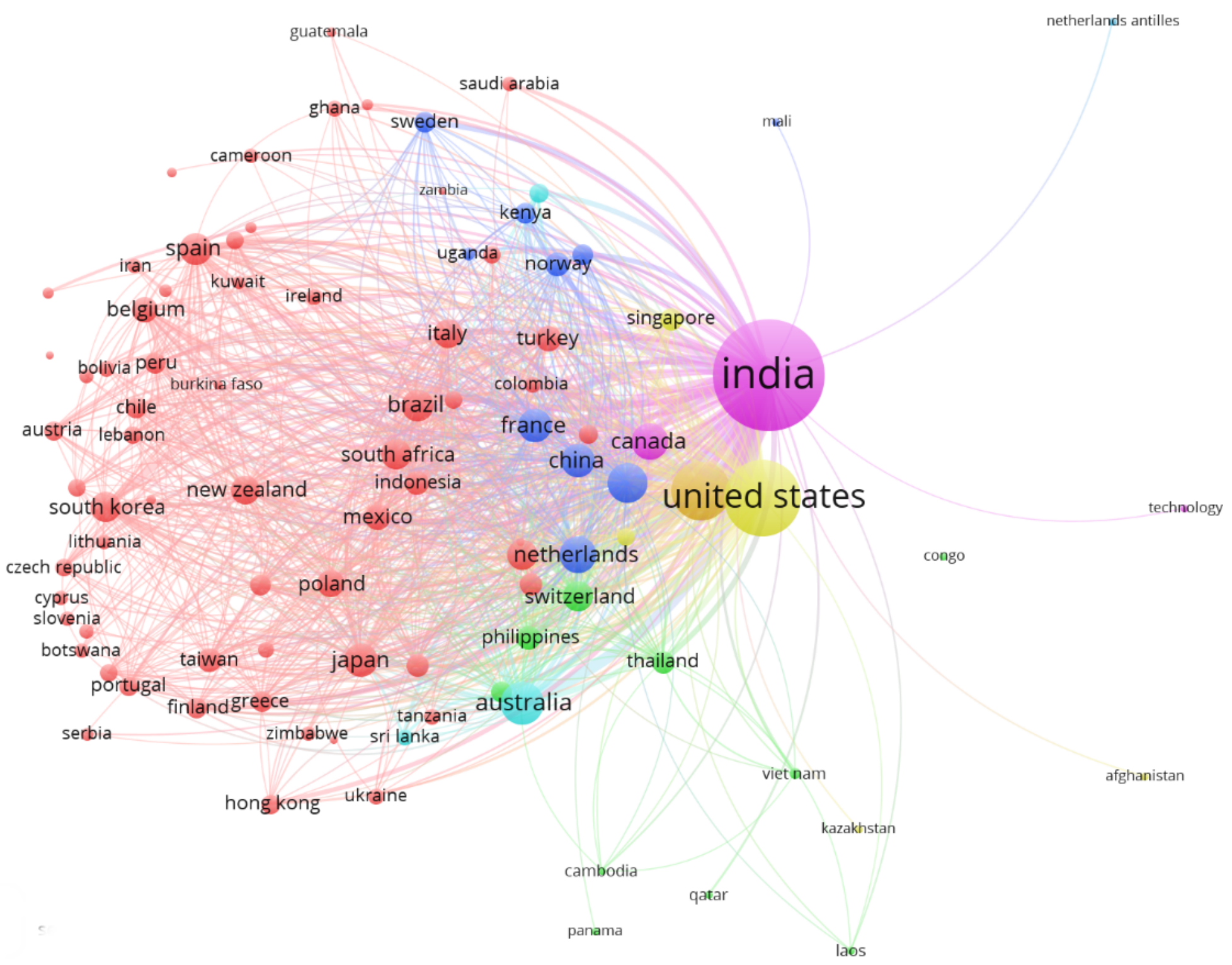

**Figure 8:** Bibliographic coupling network of India research publications

**Table 6:** Link Strength of Bibliographic coupling network of India

| Country | Bibliographic Strength |
|---|---|
| USA | 274721 |
| UK | 156947 |
| Australia | 85469 |
| Germany | 76553 |
| Canada | 61951 |
| Netherlands | 61947 |
| France | 54974 |
| China | 54851 |
| Japan | 51262 |
| Brazil | 50488 |

When we strength table of India's data then it is clear that Pakistan is not among top ten countries which are bibliographic coupled with India. USA is among top. And Brazil comes at last of the top ten list. Here important point is none of these countries match with India with respect to Cultural back ground. From this list of countries, only China and Japan are from Asia and all of other countries belongs to other continents. Despite of this factor, these countries are bibliographic coupled with Indian social science research community. Having similar social issues and resemblance in cultural back ground, Pakistan is not bibliographic coupled with India to that extend.

### 4.6 Key Word Co-occurrence analysis

Having common social issues faced by both research communities in India and Pakistan, our aim is to study whether researchers are doing work in same course or not. We can identify different areas of research by extracting author's defined keywords, by applying clustering techniques to highlight important words or areas discussed in research publication (Hofmann, 1999). To understand important topics or related field's discussion, it is important to extract keywords and then make clusters of these keywords based on their resemblance. These related keywords help to identify one topic or field. For example, if a cluster contains keywords of computer science related topics like routing algorithm, data mining then we can say that related field is Computer Science or Information Technology etc.  If problem sets of different authors are related then there is a high probability of keywords co-occurrences. We have extracted author's defined key words from available data set. And clustering is applied to identify research topics in both communities.

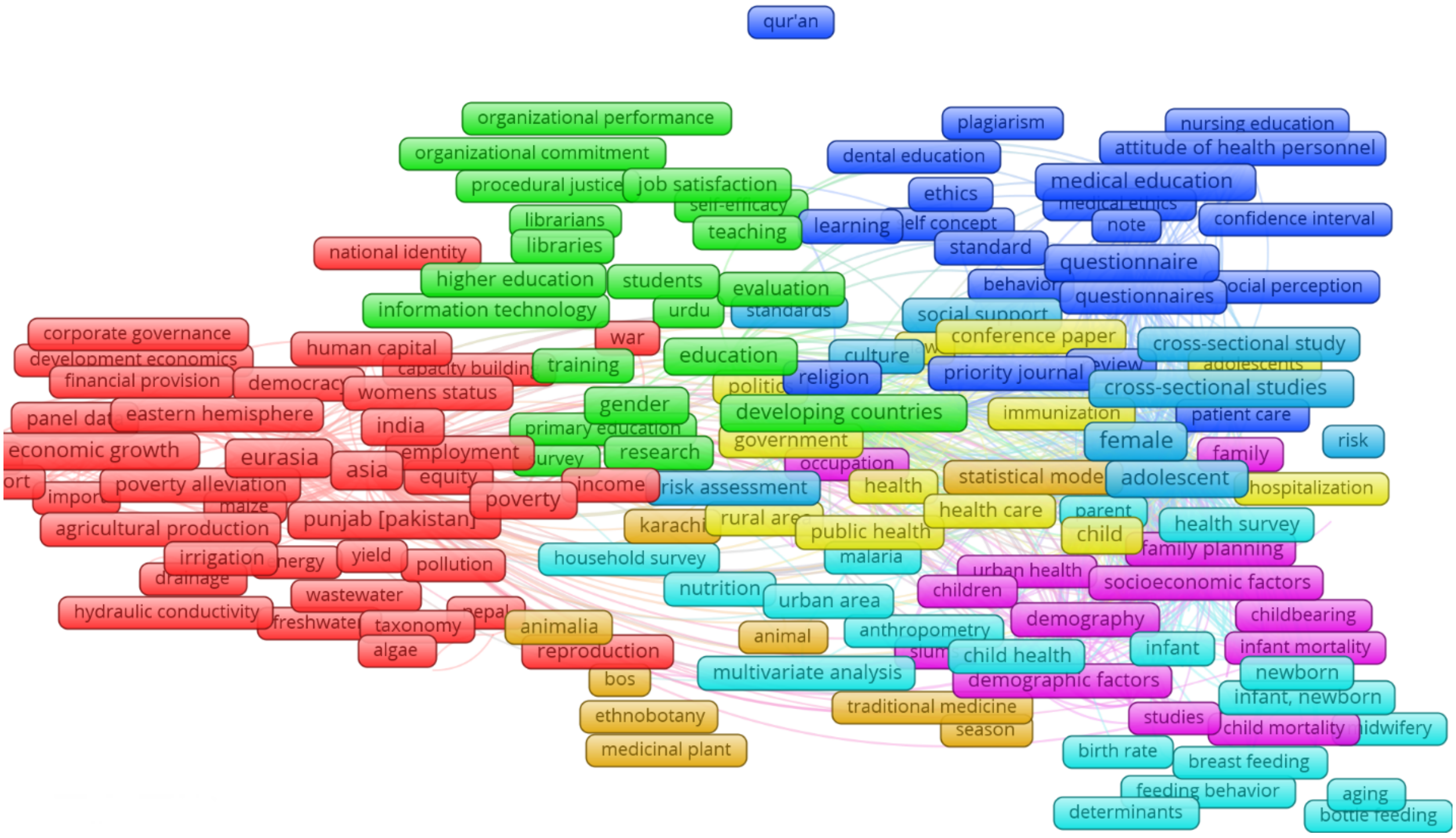

**Figure 9:** Key words Co-occurrence network defined by Pakistani authors

**Table 7:** Key words Co-occurrence defined by Pakistani authors

| **Keyword** | **Co-occurrence** |
|---|---|
| Economic growth | 89 |
| Poverty | 50 |
| Developing countries | 46 |
| Gender | 46 |
| Education | 45 |
| Job Satisfaction | 32 |
| Co integration | 31 |
| Terrorism | 25 |
| Libraries | 24 |
| Human Development | 24 |
| Foreign direct investment | 23 |

Authors from Pakistan research communities are talking more about social issues i.e. poverty, gender issues, terrorism etc. From keywords, this is obvious that authors are more focusing on social values and related areas. There are 7 prominent clusters in the given picture. Keywords with human capital, poverty alleviation, corporate governance, democracy, energy etc. are related field and we can say that here authors are discussing issue of economy or related field is economy. Similarly keywords of public health, rural areas, health care, health, immunization etc. are in cluster indicating the related of health issue. Keywords related to educations are in one cluster like higher education, Information technology, libraries, primary education, teaching etc. These keywords and clusters of these keywords are depicting clear story of ongoing research areas in the field of social science. We can also see some of the very common social issues under discussions like childbearing, hospitalization, agriculture production, urban health, infant mortality are also under discussion by Pakistani authors. The link strength is displaying top 10 keywords used by researchers from Pakistan social science community. Here most of research areas are related to social issues faced by lay man or have a direct impact on society.

Now let’s talk about clusters from Indian keyword co-occurrences.

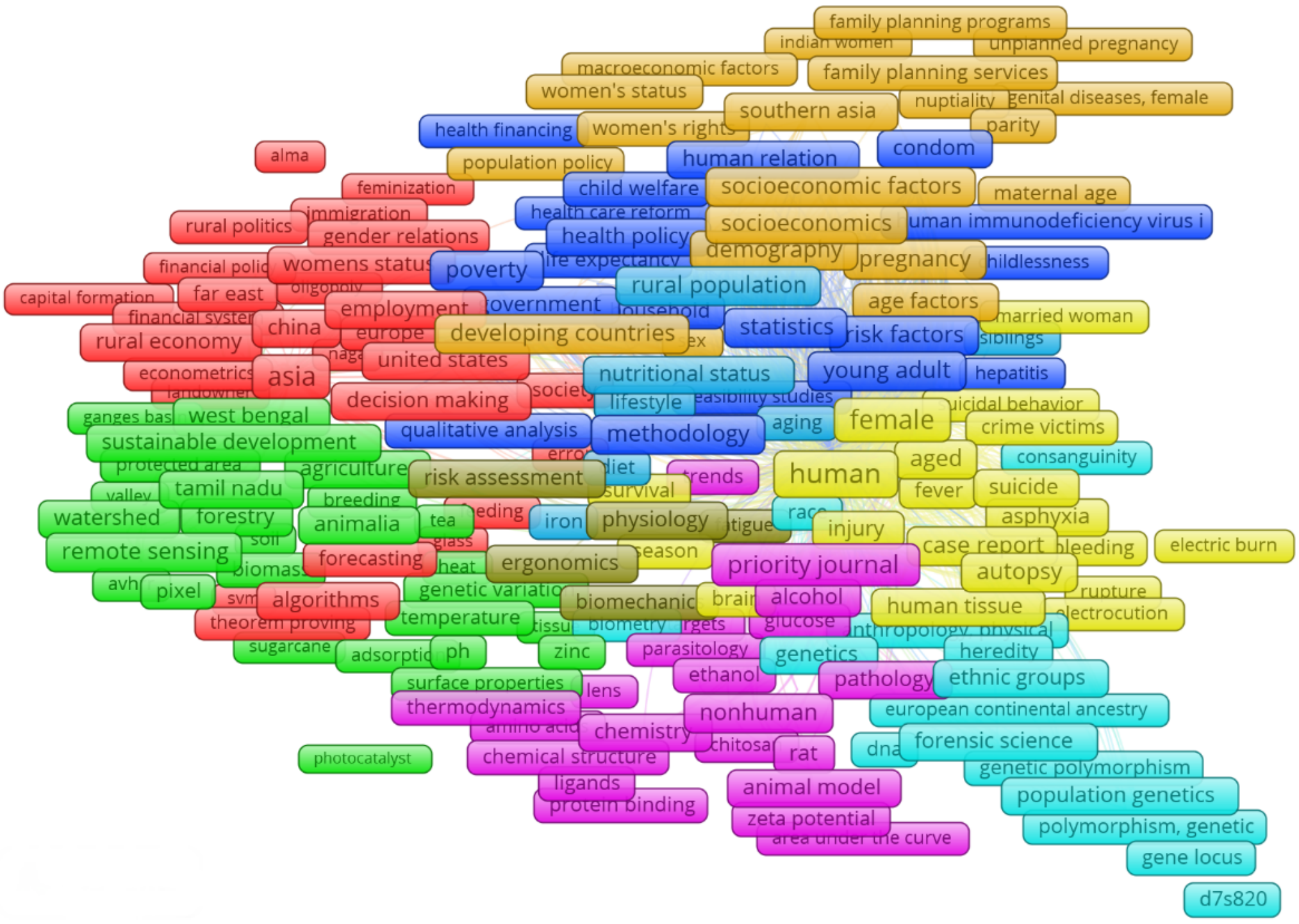

**Figure 10:** Key words Co-occurrence network defined by Indian authors

**Table 8:** Key words Co-occurrence defined by Indian authors

| Keyword | Co-occurrence |
|---|---|
| Remote sensing | 168 |
| Gender | 163 |
| Poverty | 130 |
| Climate Change | 128 |
| Education | 119 |
| Culture | 115 |
| Developing countries | 112 |
| Forensic Science | 99 |
| Sustainability | 97 |
| Developing countries | 98 |
| Agriculture | 82 |

From cluster of keywords, it is clear that authors from Indian research communities are also talking about their social issues but they are also talking about the latest educational trends within social science domain. In given cluster, keywords from education are Optical properties, nanoparticles, forensic sciences, bibliometrics, citation analysis etc. Here are top 10 keywords used by Indian Authors. From here, we can analyse that although both countries are discussing common issues but they are more likely to discuss issues faced by their own countries. For example the issue of terrorism is being discussed by Pakistani authors more likely than by Indian authors. And issue of Climate change and sustainability is more discussed by Indian research community as compared to Pakistan research community.

### 4.7 Discussion

In this section, we will summarize all of the above listed metrics. We have identified co-authorship network of both research communities with respect to the countries. Our Objective from this analysis was to see how both communities are interacting with each other and with authors of other countries. Link strength of each country has calculated and top ten collaborating countries are listed with respect to link strength. Then we have discussed about citation pattern of Pakistan and India with respect to countries and besides this, citation source pattern is also measured and identified. Quality of research publications were discussed and we compared quality of both country's publication with SJR quartile. Network maps for bibliographic coupling are also constructed and we have seen how different countries are bibliographic coupled based on strength metric. Keywords are identified for both research communities and for this purpose, we have made clusters based on author's defined keywords.

## 5. Conclusions

Based on many of the common social issues and other similarities in both communities, our hypothesis was that social science research communities from India and Pakistan are working in the same pattern. The collaboration between both research communities of India and Pakistan will be stronger than other countries. By seeing network graph, research from both communities are not collaborating with each other to that extent as compared to other countries. If we see network graph of co-authorship, then we can see that Pakistan is interacting with USA and UK. While India comes at 9$^{th}$ number in collaboration among authors. After applying clustering, it is identified that collaboration with India is not direct but authors from other countries are also involved in this research. Both countries are not present in same clusters. Similarly India is also working with countries from Europe and these countries exist in same clusters. In network graph of Pakistan's data, Muslim countries are also prominent for example Saudi Arabia and Malaysia. These countries also belong to Asia. But collaboration of these countries is weak with India. Bibliographic coupling and citation of work is also on the same pattern.

The quality of India's research publications is high as compared to publication from Pakistan. India is publishing in Q1 rank Journal with 0.46% time more than of Pakistan's publication. Other than that, publications from India as in Q2 rank journals are more in number. While work of Pakistan's authors are publishing in Q4 rank journal with factor of 49.8%. Which means around 50% of Pakistan's research publications are publishing in Q4 rank Journals. From author's defined keywords analysis for both communities, it is also clear that both communities are more likely to discuss local issues instead of common social issues. For example, Pakistan's authors are discussing about Terrorism, poverty Education, Job satisfaction while focus of Indian authors is on Remote sensing, Gender, Poverty, Climate Change, Education etc. It is important to see here that some common issues e.g Gender, Education etc. are also being discussed by authors of both communities. From all listed metrics, keywords co-occurrence is the only one metric where authors are somehow seems to work on some of common area. But clusters are entirely different for both countries.

Researchers from both communities are more likely to work with other developed countries like USA, UK, and Australia despite of cultural differences. Although research collaboration among both communities can be observed in network graph but authors from both countries are also involved in that particular collaboration. Researchers from both countries are more focused in working with developed countries despite of cultural dissimilarity.

In this work, we have used data from Scopus of social science field for Pakistan and India. And after that we applied clustering method to find collaboration with respect to co-authorship, citation, bibliographic coupling etc. In future, we will apply other mining algorithms to train our data. Then based on this training, we will analyze behavior and predict the future collaboration for both communities in field of social science. This data will also be used to see behavior of network graphs of other fields. Then we will check accuracy and of system based on this training data. For future work, we are developing an intelligent data repository that extracts scientific literature where the cited work has considerable contribution. And then comparison will be made on all of sub fields of social sciences. It will help in identifying sub fields where authors from both countries are collaborating extensively.